\documentclass[numberedappendix]{emulateapj}
\usepackage{epsfig, natbib}

\newcommand{\msun}{M_{\odot}}

\newcommand{\krho}{k_{\rho}}

\newcommand{\rct}{\tilde{R}}
\newcommand{\Lt}{\tilde{L}}

\newcommand{\kt}{k_{\rm T}}
\newcommand{\mbe}{M_{\rm BE}}

\newcommand{\mh}{m_{\rm H}}

\newcommand{\mr}{m_{\rm R}}
\newcommand{\calc}{\mathcal{C}}
\newcommand{\sigmat}{\sigma_{\rm T}}
\newcommand{\ltsim}{\protect\raisebox{-0.5ex}{$\:\stackrel{\textstyle <}
	{\sim}\:$}}

\newcommand{\jcap}{JCAP}

\begin{document}

\slugcomment{ApJ in press}

\title{On the Origin of Stellar Masses}

\shorttitle{On the Origin of Stellar Masses}
\shortauthors{Krumholz}

\author{Mark R. Krumholz}
\affil{Department of Astronomy, University of California, Santa Cruz, CA 95064}

\begin{abstract}
It has been a longstanding problem to determine, as far as possible, the characteristic masses of stars in terms of fundamental constants; the almost complete invariance of this mass as a function of the star-forming environment suggests that this should be possible. Here I provide such a calculation. The typical stellar mass is set by the characteristic fragment mass in a star-forming cloud, which depends on the cloud's density and temperature structure. Except in the very early universe, the latter is determined mainly by the radiation released as matter falls onto seed protostars. The energy yield from this process is ultimately set by the properties of deuterium burning in protostellar cores, which determines the stars' radii. I show that it is possible to combine these considerations to compute a characteristic stellar mass almost entirely in terms of fundamental constants, with an extremely weak residual dependence on the interstellar pressure and metallicity. This result not only explains the invariance of stellar masses, it resolves a second mystery: why fragmentation of a cold, low-density interstellar cloud, a process with no obvious dependence on the properties of nuclear reactions, happens to select a stellar mass scale such that stellar cores can ignite hydrogen. Finally, the weak residual dependence on the interstellar pressure and metallicity may explain recent observational hints of a smaller characteristic mass in the high pressure, high metallicity cores of giant elliptical galaxies.
\end{abstract}


\section{Introduction}

The question of the origin of the stellar mass scale dates back to the time of Eddington and Jeans. Modern observations reveal that the median stellar mass remains unchanged in star forming environments that vary by orders of magnitude in density, pressure, metal content, and other variables \citep{bastian10a}, which suggests that this mass must be set mostly by fundamental constants. A significant advance toward solving the problem has come from the realization that the characteristic fragmentation mass of interstellar clouds is determined primarily by their temperature structures \citep{spaans00a, larson05a}. In retrospect this is not surprising, and can in fact be deduced simply from dimensional analysis. An isothermal, turbulent, self-graviting, magnetized gas is fully characterized by three dimensionless numbers (e.g.\ the ratio of gas to magnetic pressure $\beta_{\rm mag}$, the Mach number $\mathcal{M}$, and the virial ratio $\alpha_{\rm vir}$), but these ratios admit a rescaling that leaves the dimensionless numbers fixed but changes the mass scale arbitrarily. (For a formal proof see the appendix of \citealt{mckee10b}). As a result, any simulation or analytic calculation of the evolution of an isothermal cloud can always be rescaled to produce objects of arbitrary mass, and the characteristic mass found in isothermal simulations depends mostly on the numerical resolution used \citep{martel06a}. For this reason, analytic theories of the initial mass function (IMF; e.g.~\citealt{padoan02a, hennebelle08b}) or simulations of star cluster formation \citep[e.g.][]{girichidis11a} that assume a purely isothermal equation of state may be able to predict the functional form of the stellar mass {\it distribution} as a function of quantities like $\mathcal{M}$, $\beta_{\rm mag}$, or $\alpha_{\rm vir}$, but they are always forced to leave the absolute mass {\it scale} as a free parameter. Thus any explanation of the characteristic stellar mass and its invariance must somehow depend on deviations from isothermality.

Given this realization, attention has therefore turned to the question of what processes determine the gas temperature structure, and what mass scale they select. One idea is that the characteristic stellar mass is set by a change from poor to strong dust-gas coupling at a density of $\sim 10^4$ cm$^{-3}$ \citep{larson05a, elmegreen08a}. However, this explanation faces the problem that the typical $\sim 10^4$ $\msun$ region of star cluster formation in the Milky Way is an order of magnitude denser than this \citep[e.g.][]{shirley03a, faundez04a, fontani05a}, yet still manages to fragment down to masses $\la \msun$. Another proposed mechanism that applies to gas near galactic centers or in ULIRGs is heating by either x-rays or cosmic rays \citep[e.g.][]{klessen07a, papadopoulos10a, hocuk10b, hocuk10a, meijerink11a}. While these effects may indeed yield deviations from the canonical IMF in extreme conditions, they do not apply to most star-forming environments. Even near the Galactic center, observations so far have yet to find any evidence for their influence on the IMF \citep[e.g.][]{brandner08a}.

Instead, the primary mechanism for determining the temperature structure of star-forming gas clouds appears to be radiation feedback produced by accretion onto the protostars themselves \citep{krumholz06b, whitehouse06a}. This is the dominant energy source in a star-forming cloud \citep{offner09a}, and at metallicities $\ga 1\%$ of the Solar value \citep{omukai10a, myers11a} and densities $\ga 10^4$ cm$^{-3}$ \citep{goldsmith01a}, which characterize almost all star-forming environments we are able to observe, this energy is well-coupled to the gas. Simulations and analytic estimates show that it changes how gas fragments \citep{krumholz07a, krumholz10a, krumholz11c, krumholz08a, bate09a, offner09a}.

The only attempt thus far to understand why radiative feedback leads to an invariant peak of the stellar IMF is that of \citet{bate09a}, who gives a scaling argument for why the density should have little effect on the IMF peak. In this paper I expand and improve this argument in several ways. First, \citeauthor{bate09a} relies on an empirically-determined mass-radius relation to set the protostellar luminosity, and it is not clear how this might vary with star-forming environment or what physics sets it. I demonstrate that the necessary relation can be obtained, at least approximately, from fundamental constants. Second, the argument in \citeauthor{bate09a} is limited to the case where the gas is optically thin. However, the optical depth of a protostellar core varies radically as a function of wavelength, and it is not clear at what wavelength the condition of optical thinness must be satisfied. Consequently, it is not clear to which, if any, star-forming environments Bate's argument can be applied. In contrast, the calculation I present here requires no assumptions about the optical depth of the star-forming region. Third, \citeauthor{bate09a} only gives a scaling argument for why the characteristic stellar mass depends little on the ambient density, but does not actually estimate what this mass scale is. I provide such an estimate in terms of fundamental constants. The argument presented here therefore significantly expands the theory first advanced by \citet{krumholz06b} and \citet{bate09a} for how radiative feedback can set the stellar mass scale.

\section{Calculation of the Characteristic Mass}

\subsection{Fragmentation of Interstellar Clouds}

Consider a region of mean pressure $P$ that begins to collapse to form a star. Because the dynamical time varies with volume density $\rho$ as $t_{\rm dyn} \approx 1/\sqrt{G\rho}$, the densest portion of a cloud has the shortest dynamical time, and rapidly collapses to form a thermal pressure-supported seed protostar. The surrounding gas may then either accrete onto the existing protostar, or it may collapse independently to form additional stars. We can idealize the fragmentation of this gas as a competition of two processes. Near the star, the gas will be heated by the star's radiation output, and this will raise its pressure and make it resistant to fragmentation. This material is therefore likely to accrete. Far from the star, the gas is colder, and it is therefore likely to fragment into new stars rather than accreting onto the existing one. The goal then is to compute the typical mass $M$ of gas that is heated to the point where it will accrete rather than fragment -- in effect to compute the protostar's thermal zone of influence. 

In order to make this estimate, I approximate that this mass $M$ has a density distribution $\rho\approx \rho_e (r/R)^{-\krho}$, where the radius $R$ is to be determined and $\rho_e = [(3-\krho)/4\pi] M/R^3$. The choice of $\krho$ is arbitrary, and I show below that it makes almost no difference. Obviously the overall assumption of spherical symmetry is not fully realistic, but this assumption enables us to perform an analytic calculation yet still capture the essential physics. The virial theorem implies that, for non-magnetized material in virial balance, the mass and radius are related via the external pressure by \citep{krumholz05c}
\begin{equation}
\label{eq:pressure}
P\approx \frac{3}{20\pi} \alpha_{\rm vir} \frac{G M^2}{R^4},
\end{equation} 
where $\alpha_{\rm vir}$ is the virial ratio. For a marginally bound object $\alpha_{\rm vir} \approx 2$, and I adopt this value throughout. One might worry that the effective pressure might be enhanced by the ram pressure of inflow. However, observations of the flows around protostars indicates that, on the scales of individual low-mass protostellar cores, the inflow velocity is at most transonic, indicating that the infall ram pressure cannot be much larger than the above estimate \citep{andre07a, kirk07a, rosolowsky08a, friesen09a, friesen10a, maruta10a}. This lack or strong shocks within cores is consistent with theoretical models of turbulent fragmentation \citep{offner08a, gong09a}. One might also worry that, if protostellar cores are sufficiently dominated by magnetic pressure, they may have $\alpha_{\rm vir} \ll 1$. In the Milky Way this does not appear to be the case \citep[e.g.][]{lada08a}, but we cannot directly rule out the possibility that cores are highly magnetic pressure-dominated in other galaxies. One should keep this caveat in mind.

\begin{figure}
\epsscale{1.18}
\plotone{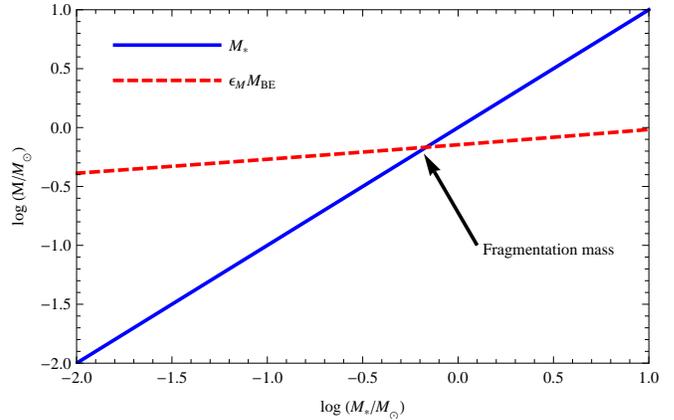}
\caption{
\label{fig:mvsm}
Stellar mass $M_*$ (blue solid line) and Bonnor-Ebert mass times efficiency $\epsilon_M M_{\rm BE}$ (red dashed line) as a function of stellar mass $M_*$. The Bonnor-Ebert mass is computed using the formalism described by equations (\ref{eq:trho}) -- (\ref{eq:psi}), using the fiducial parameters $\krho=1.5$, $\delta=1$, $\beta=2$, $n=3/2$, $\epsilon_M = 1/2$, $\epsilon_L = 3/4$, and computed at a pressure $P/k_B=10^7$ K cm$^{-3}$. Note that $\epsilon_M M_{\rm BE} \gg M_*$ for small $M_*$, but that this reverses at large $M_*$. The intersection of the two lines gives the estimated stellar mass set by fragmentation, as indicated by the arrow.\\
}
\end{figure}

The minimum mass that is capable of undergoing gravitational instability and collapsing to form a second, separate protostar is the Bonnor-Ebert mass, $\mbe = 1.18 c_s^3/\sqrt{G^3\rho}$, where $c_s$ is the gas sound speed. Numerical simulations including magnetic fields show that they do not significantly alter this characteristic fragment mass \citep{padoan07a}. Thus if we consider successively larger spheres surrounding the first protostar, the smallest such sphere that contains enough mass to be capable of fragmenting and forming another star has a mass $\mbe$. The material interior to that will therefore have to accrete onto the first star (or be ejected by its outflow). We may therefore think of the time-evolution of the system as follows. The instant after a protostar forms, its mass $M_*$ is very small. In contrast, the Bonnor-Ebert mass in the gas around it is
\begin{equation}
M_{\rm BE} = 1.18 \sqrt{\left(\frac{k_B T_e}{\mu_{\rm H_2} \mh G}\right)^3 \frac{1}{\rho_e}},
\end{equation}
and this is much larger than $M_*$. Consequently, although the protostar is small, it has a much larger reservoir of gas around it that is too warm to fragment, and will instead accrete. As the star gains mass, both $M_*$ and $M_{\rm BE}$ rise, but $M_{\rm BE}$ rises much more slowly -- the star is consuming mass faster than its reservoir is growing. Once the mass accreted onto the star is equal to the entire mass of the heated reservoir, fragmentation becomes likely and the star will stop growing. This condition provides our estimate for the characteristic stellar mass:
\begin{equation}
\label{eq:mch1}
M_* = \epsilon_M M_{\rm BE} = 1.18 \epsilon_M \sqrt{\left(\frac{k_B T_e}{\mu_{\rm H_2} \mh G}\right)^3 \frac{1}{\rho_e}},
\end{equation}
where $\epsilon_M\approx 1/2$ \citep{matzner00a, alves07a, enoch08a} is the fraction of the mass that has collapsed ({\it not} the fraction of the entire available mass reservoir) that has been incorporated into the star rather than being ejected by the protostellar outflow, $T_e$ is the gas temperature at the edge of the region that will form the star, $\mh$ is the hydrogen mass, and $\mu_{\rm H_2}=2.33$ is the mean particle mass (in units of $\mh$) for a molecular hydrogen-dominated gas of Solar composition. Figure \ref{fig:mvsm} illustrates this procedure graphically: at small $M_*$, the available reservoir mass $\epsilon_M M_{\rm BE} \gg M_*$, but as $M_*$ rises, eventually the two become equal and fragmentation sets in.

The temperature $T_e$ is set by the radiation of the central star. If this star has luminosity $L$, this is well-approximated by \citep{chakrabarti05a, chakrabarti08a, myers11a}
\begin{eqnarray}
T_e^\gamma & = & \left(\frac{L/M}{4\sigma_{\rm SB} \Lt}\right)^{\krho-1+\beta\kt}
\left[\frac{(3-\krho) \delta \kappa_0}{4(\krho-1) T_0^\beta}\right]^{4\kt-2}
\nonumber
\\
& &
{} \times \left(\frac{M}{\pi R^2}\right)^{(4+\beta)\kt+\krho-3},
\label{eq:trho}
\end{eqnarray}
where $\gamma = 2\beta+4(\krho-1)$. Here the dust opacity is taken to follow a wavelength-dependence $\kappa_\lambda = \delta \kappa_0(\lambda_0/\lambda)^\beta$, where $\delta$ is a dimensionless number and we arbitrarily define $\kappa_0 = 0.27$ cm$^2$ g$^{-1}$, $\lambda_0 = 100$ $\mu$m, and $T_0 = hc/\lambda_0 k_B = 144$ K. With this parameterization $\delta\approx 1$ and $\beta\approx 2$ for Milky Way dust \citep{weingartner01a}. Dust coagulation can alter $\delta$ by factors of a few and $\beta$ by a few tenths \citep{ossenkopf94a}, but I show below that the results are extremely insensitive to these variations of this magnitude. The index $\kt$ and the dimensionless constant $\Lt$ in turn are given by
\begin{eqnarray}
\label{eq:kt}
\kt & \approx & \frac{0.48 \krho^{0.005}}{\rct^{0.02 \krho^{1.09}}} + \frac{0.1 \krho^{5.5}}{\rct^{0.7\krho^{1.09}}} \\
\label{eq:lt}
\Lt & \approx & 1.6 \rct^{0.1} \\
\label{eq:rt}
\rct & = & \left\{\frac{(L/M)(M/\pi R^2)^{(4+\beta)/\beta}}{4\sigma_{\rm SB}\Lt}\left[\frac{(3-\krho)\delta\kappa_0}{4(\krho-1)T_0^\beta}\right]^{4/\beta}\right\}^{-\beta/\gamma}.
\end{eqnarray}
Note that this calculation assumes that the only significant source of luminosity is the single accreting protostar at the center, which is true only if the thermal zones of influence of different stars do not overlap. Numerical simulations and analytic calculations by \citet{krumholz11c} show that this is a good approximation as long as the star formation rate in a protocluster where stars are forming is $\la 10\%$ of the mass per free-fall time. All observed star-forming regions, both within and outside the Galaxy, obey this constraint \citep{krumholz07e, evans09a}. The calculation also assumes that the heated region is spherically symmetric. Numerical simulations show that that this is not a bad approximation, since the highly diffusive nature of the radiation-matter interaction tends to produce fairly round heated regions even in the presence of asymmetric features such as accretion disks \citep[e.g.][]{offner09a, bate09a}.

The luminosity of young low mass stars is dominated by accretion, so $L$ will be proportional to the accretion rate. This is $\dot{M}_* \approx \epsilon_M M/t_{\rm dyn}$, where $t_{\rm dyn} \approx 1/\sqrt{G\overline{\rho}} = \sqrt{(3-\krho)/(3 G \rho_e)}$ is the dynamical time in the collapsing region. For this accretion rate, the corresponding luminosity is
\begin{equation}
\label{eq:lum}
L = \epsilon_L \frac{G M_*}{R_*} \dot{M}_* = \epsilon_L \epsilon_M \sqrt{\frac{3 G \rho_e}{3-\krho}} M \psi
\end{equation}
where $M_*$ and $R_*$ are the stellar mass and radius, $\psi \equiv G M_*/R_*$ is the energy yield per unit mass for accreted matter, and $\epsilon_L\approx 3/4$ is the fraction of the accretion power that goes into light rather than into driving an outflow \citep{mckee03a}.\footnote{Note that $\epsilon_L$ can also be reduced by episodic accretion that delivers some fraction of the final mass in short-duration bursts; however, comparisons with the observed protostellar luminosity function suggest that the reduction in luminosity during the non-burst phase is modest, only $\sim 25\%$ \citep{offner11a}, and I show below that the characteristic mass depends fairly weakly on $\epsilon_L$. Thus I do not attempt to include this effect.} Equations (\ref{eq:pressure}), (\ref{eq:mch1}), (\ref{eq:trho}), and (\ref{eq:lum}), together with $\psi$, fully determine $M_*$. Computing $\psi$ is therefore the next task.

\subsection{The Stellar Mass-Radius Relation from Fundamental Physics}

The energy yield from accretion $\psi$ is dictated by a number of factors, but the single most important one is deuterium burning. For almost all stars this sets during accretion, and it forces the stellar core to a nearly fixed central temperature. I approximate D-burning stars as $n=3/2$ polytropes, although different values of $n$ produce qualitatively identical results. Assuming the stellar core is fully ionized and dominated by ideal gas pressure, this implies that $\psi = T_n (k_B T_c/\mu_i \mh)$, where $T_c$ is the central temperature, $\mu_i=0.61$ is the mean mass per particle for a fully ionized gas of Solar composition, and $T_n$ is a dimensionless number that depends on the polytropic index; for $n=3/2$, $T_n = 1.86$ \citep{chandrasekhar39a}. 

I estimate the equilibrium central temperature $T_c$ following the formalism of \citet{adams08a}, in which a star of mass $M_*$ and radius $R_*$ is approximated as a polytrope of index $n$, for which the density distribution follows $\rho(\xi)=\rho_c f^n(\xi)$, where $\rho_c$ is the central density, $\xi = r/R_*$ is the dimensionless radius, and $f(\xi)$ is the solution to the Lane-Emden equation
\begin{equation}
\frac{d}{d\xi} \left(\xi^2\frac{df}{d\xi}\right) + \xi^2 f^n = 0,
\end{equation}
with the boundary conditions $f(0) = 1$, $f'(0) = 0$. Assuming the star is dominated by ideal gas pressure, the temperature is then $T = T_c f(\xi)$. From this solution, the dimensionless decay rate for the temperature as a function of radius is $\beta = \xi_1^{-1}$, where $\xi_1$ is defined by the condition $f(\xi_1)=e^{-1}$; the dimensionless mass is
\begin{equation}
\mu_0 = \int_0^{\xi_*} \xi^2 f^n(\xi)\,d\xi,
\end{equation}
where $\xi_*$ is defined by the condition $f(\xi_*) = 0$.

The next step in the calculation is to estimate the rate of nuclear energy generation per unit volume using the Laplace approximation \citep{fowler75a}, which gives
\begin{equation}
\epsilon = \calc \rho^2 \Theta^2 \exp(-3\Theta),
\end{equation}
where $\calc$ is a constant that depends on the properties and abundance of the reactants, $\Theta =  (E_G/4k_B T)^{1/3}$, and $E_G = (\pi \alpha Z_1 Z_2)^2 2 \mr c^2$ is the Gamow energy for the reaction. Here $Z_1$ and $Z_2$ are the charges on the two reacting nuclei, $\mr$ is their reduced mass, and $\alpha$ is the fine structure constant. Comparison with detailed calculations of nuclear reaction rates \citep{fowler75a} shows that this is an excellent approximation for the D burning reaction with which we are concerned (discussed in more detail below) as long as the temperature is $\ll 10^9$ K. Note that, if we define $\Theta_c = (E_G/4 k_B T_c)^{1/3}$ and $\Theta = \Theta_c f(\xi)^{-1/3}$, the rate of nuclear energy integrated over the stellar volume is $L_* = \calc 4\pi R^3 \rho_c^2 I(\Theta_c)$, where
\begin{equation}
I(\Theta_c) = \int_0^{\xi_*} f^{2n} \xi^2 \Theta^2 \exp(-3\Theta)\, d\xi.
\end{equation}

Given these approximations, \citet{adams08a} shows (his equation 25) that the conditions of thermal and hydrostatic balance within the star require that the central temperature roughly obey
\begin{equation}
\label{eq:thetac}
I(\Theta_c) \Theta_c^{-8} = \frac{2^{12}\pi^5}{45} \frac{1}{\beta \kappa_c \calc E_G^3 \hbar^3 c^2} \left(\frac{M_*}{\mu_0}\right)^4 \left(\frac{G \mu_i \mh}{n+1}\right)^7,
\end{equation}
assuming the star is fully ionized so the mean particle mass is $\mu_i \mh$. Here $\kappa_c$ is the opacity at the center of the star, and which we approximate as dominated by Thompson scattering, so $\kappa_c = \sigmat/\mh (1+X)/2$, where $\sigmat$ is the Thompson cross section $X$ is the hydrogen mass fraction. The results are quite insensitive to changes in $\kappa_c$ by factors of a few, or even tens. Note that equation (\ref{eq:thetac}) is derived assuming that energy transport is by radiation rather than convection, which is true only in part of the star during D burning. However, below I compare the value of $\psi$ derived from this assumption to the results of a detailed numerical model that includes convection, and show that convection alters $\psi$ by at most a factor of a few. 

In order to solve equation (\ref{eq:thetac}) for the central temperature $\Theta_c$, the final necessary step is to compute the reaction constant $\calc$, defined by
\begin{equation}
\calc =  \frac{\langle \sigma v\rangle}{\Theta^2 \exp(-3\Theta)} \left(\frac{\langle \Delta E \rangle}{\mu_1\mu_2 \mh^2}\right),
%\frac{8 \langle \Delta E\rangle S(E_0)}{\sqrt{3}\pi \alpha \mu_1 \mu_2 \mh^2 Z_1 Z_2 \mr c},
\end{equation}
where $\sigma v$ is the velocity times cross section for the reaction,  $\Delta E$ is the net energy released per reaction, $\mu_1$ and $\mu_2$ are the mean mass (in units of $\mh$) per reactant of species $1$ and $2$, and the angle brackets indicate averages over the Maxwellian velocity distribution of reacting particles. Note that, since $\langle \sigma v\rangle$ depends on the temperature as $\Theta^2 \exp(-3\Theta)$ in the Laplace approximation, this quantity is temperature-independent. 

At the $\sim 10^6$ K temperatures typical of deuterium-burning stars, the dominant D burning reaction chain by a large margin is \citep{stahler80a}
\begin{eqnarray}
{}^2_1{\rm D} + {}^1_1{\rm H} & \rightarrow & {}^3_2{\rm He} \\
2\, {}^3_2{\rm He} & \rightarrow & {}^4_2{\rm He} + 2\, {}_1^1{\rm H},
\end{eqnarray}
which yields $\langle \Delta E \rangle = 12.6$ MeV per D burned, with negligible neutrino losses. The first reaction is the rate-limiting step. Its Gamow energy is $E_G = (4/3) \pi^2 \alpha^2 \mh c^2 = 0.66$ MeV, and the mean masses per particle for the two reactant species are $\mu_1 = \mu_{\rm H} = \mh/X$ and $\mu_2 = \mu_{\rm D} = \mu_{\rm H}/[{\rm D}/{\rm H}]$, where $[{\rm D}/{\rm H}]$ is the abundance ratio of deuterium relative to hydrogen. For interstellar gas, and thus gas at the onset of D burning, $[{\rm D}/{\rm H}] \approx 2\times 10^{-5}$ \citep{stahler80a}; the ratio will decline with time during the main D burning phase, but this does not affect $\Theta_c$ significantly until essentially all the D is burned, as I show below. For $X=0.71$, and the value of $\langle\sigma v\rangle$ tabulated by \citet{fowler75a}, I find $\calc=2.1\times 10^{17} ([{\rm D}/{\rm H}]/2\times 10^{-5})$ cm$^5$ s$^{-3}$ g$^{-1}$.

\begin{figure}
\epsscale{1.18}
\plotone{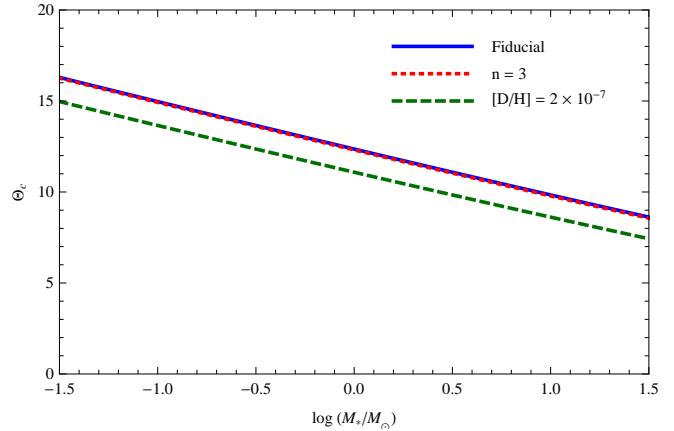}
\caption{
\label{fig:thetac}
Dimensionless central temperature $\Theta_c$ as a function of stellar mass $M_*$, computed from equation (\ref{eq:thetac}). The solid blue line shows the fiducial case of ($n=3/2$ polytrope, $[{\rm D}/{\rm H}]=2\times 10^{-5}$). The red dotted line shows $n=3$, and the green dashed line shows $[{\rm D}/{\rm H}]=2\times 10^{-7}$.\\
}
\end{figure}

\begin{figure}
\epsscale{1.18}
\plotone{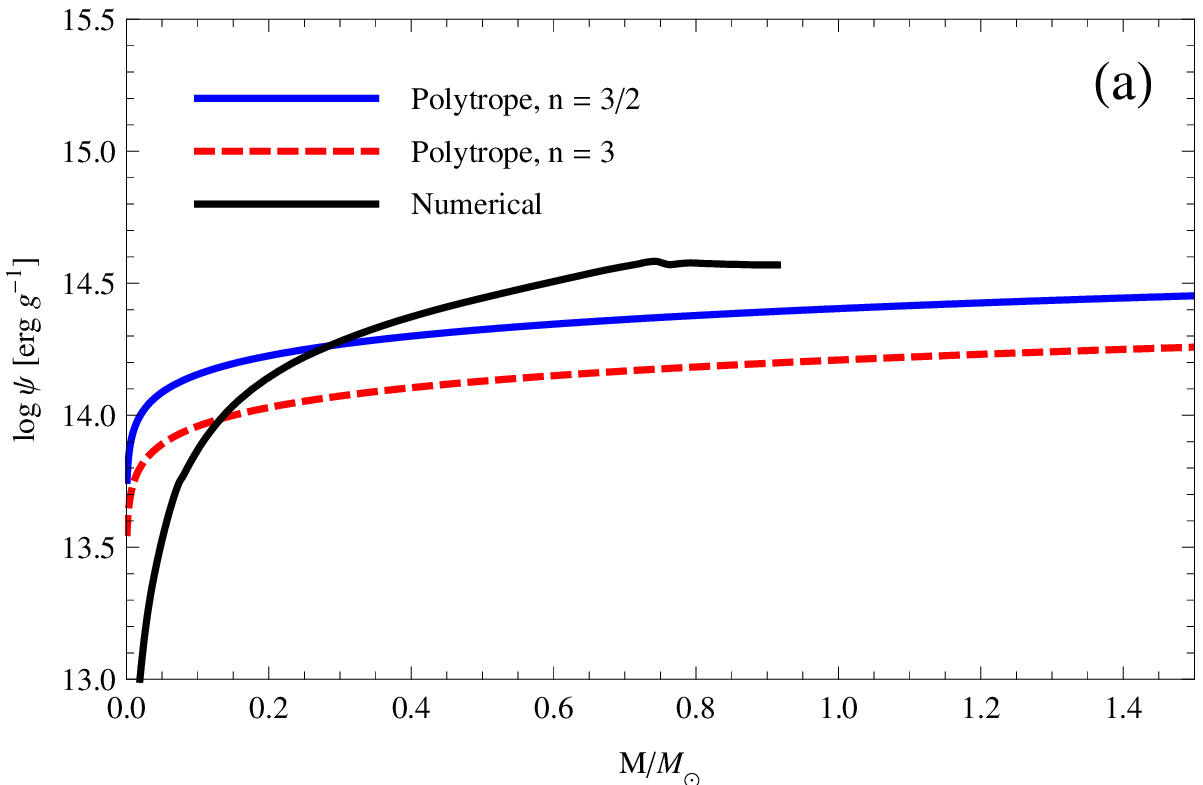}
\plotone{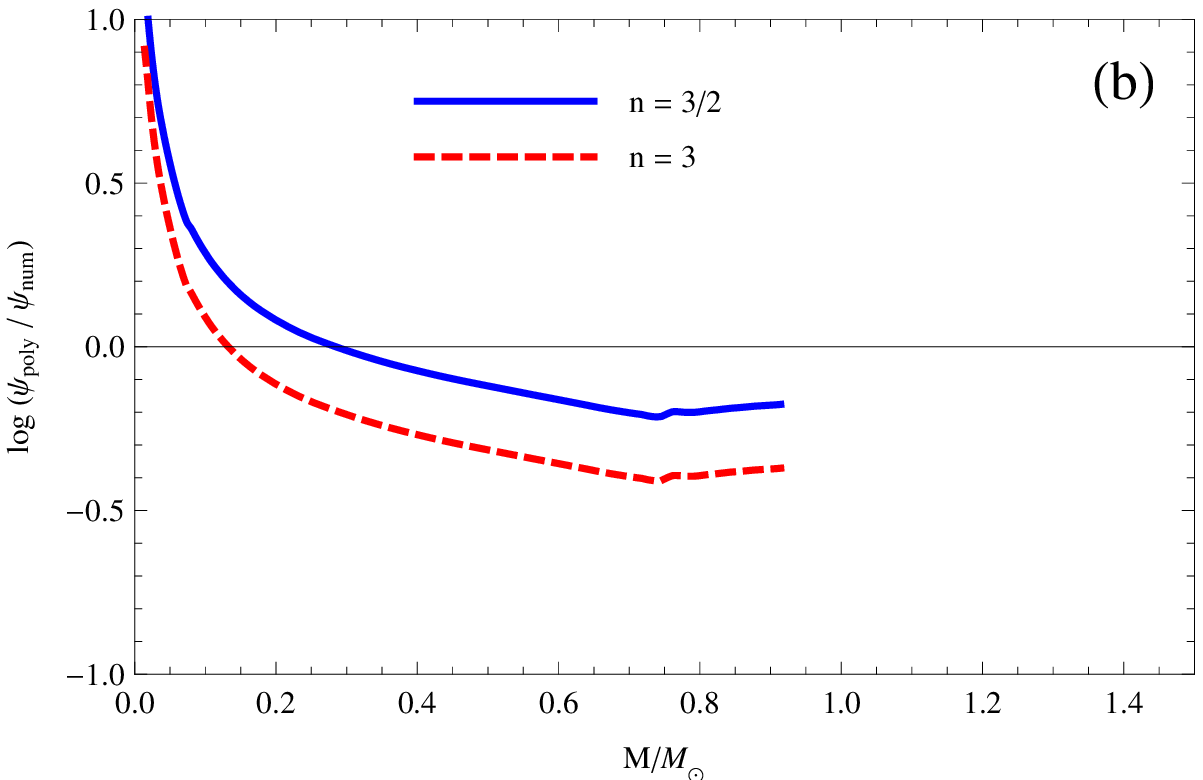}
\caption{
\label{fig:psiplot}
(a) Energy per unit mass $\psi$ released by accretion onto a protostar of mass $M$, from equation (\ref{eq:psi}), for $n=3/2$ (solid blue line) and $n=3$ (red dashed line) polytropes; the solid black line shows the results of a detailed stellar structure calculation (model mC5H of \citealt{hosokawa11a}). (b) Ratio of the polytropic estimate of $\psi$ to the numerically-determined value, for polytropic indices $n=3/2$ (solid blue) and $n=3$ (dashed red). Note that the \citeauthor{hosokawa11a} models are initialized to a mass of $0.01$ $\msun$, and that the results are highly sensitive to the choice of initial radius until the stellar mass reaches several times this value. Thus one only should take the \citeauthor{hosokawa11a}\ models seriously at masses $\ga 0.05$ $\msun$.\\
}
\end{figure}

Numerically solving equation (\ref{eq:thetac}) using this value of $\calc$ gives the results shown in Figure \ref{fig:thetac}. We see that $\Theta_c$ is almost completely insensitive to changes in the polytropic index, and varies by only tens of percent as $M_*$ or $[{\rm D}/{\rm H}]$ vary by orders of magnitude. The central temperature is $T_c \approx E_G / 4 k_B \Theta_c^3$. Thus we have
\begin{equation}
\label{eq:psi}
\psi = \left(\frac{T_n}{4\Theta_c^3}\right) \frac{E_G}{\mu_i \mh}.
\end{equation}
For $M_* = \msun$, I find $T_c = 1.0\times 10^6$ K, and $\psi = 2.5\times 10^{14}$ erg g$^{-1}$, which agrees to within a factor of $\sim 2$ with the results of detailed stellar structure models \citep{stahler80a, hosokawa11a}. Figure \ref{fig:psiplot} shows a more detailed comparison. As the plot shows, the polytropic estimate agrees with the numerical result to better than half a dex at all masses $\ga 0.05$ $\msun$, and to better than a dex at all masses $\ga 0.01$ $\msun$. I show below that errors of this magnitude have very little effect on the final result.

\subsection{The Characteristic Stellar Mass}

Equation (\ref{eq:psi}) completes the system formed by equations (\ref{eq:pressure}), (\ref{eq:mch1}), (\ref{eq:trho}), and (\ref{eq:lum}), and uniquely specifies $M$ and $M_*$. Before proceeding with a numerical solution, however, one can gain considerable insight from an approximate analytic solution. 
%Note that, across a broad range of parameter space, $\kt \approx 0.5$ and $\Lt\approx 1$. 
Notice that equations (\ref{eq:kt}) -- (\ref{eq:rt}) imply that $\kt \approx 0.5$ and $\Lt\approx 1$ as long as $\rct \ltsim 1/2$. The quantity $\rct$ represents the ratio of the radius of the dust photosphere to the core radius, and this will be $\ll 1$ as long as the core is opaque enough that stellar photons escape primarily by diffusing to frequencies where the core optical depth is $\sim 1$ rather than by diffusing out of the core in space while remaining at frequencies where the core is optically thick. This is the case for almost all of the parameter space relevant to star formation, as can be verified readily from a numerical solution I give below. For the purposes of analytic approximation, therefore, it is reasonable to adopt $\kt \approx 0.5$ and $\Lt\approx 1$. Similarly, $\Theta_c$ is nearly independent of the stellar mass. If one takes these approximate values as exact, assumes $\Theta_c$ is mass-independent, and further takes $\beta = 2$ (as expected for most dust models) and $\krho = 1.5$, then it is possible to solve the system of equations analytically. After some manipulation, the result is
\begin{eqnarray}
M_* & = &
\mh \left(\frac{1.18^{64} 2^{69} 5^{21}}{3^{17} \pi^7}\right)^{1/54}
\left(\frac{T_n^4 \epsilon_L^4 \epsilon_M^{13}}{\mu_{\rm H_2}^{16} \mu_i^4}\right)^{1/9}
\nonumber
\\
& & \qquad{} \times 
\left(\frac{\alpha^{16}}{\alpha_G^{25}}\right)^{1/18} 
\Theta_c^{-4/3} \left(\frac{P}{P_P}\right)^{-1/18}
\label{eq:mstar}
%\left(\frac{\alpha^{16} c^5 \mh^{22}}{\alpha_G^{27} \Theta_c^{24} h^3}\right)^{1/18}
%\left(\frac{\alpha^{16} c^5 \mh^{4}}{\alpha_G^{27} \Theta_c^{24} h^3}\right)^{1/18} \mh
%P^{-1/18}
%\left(\frac{1.18^{64} 5^{21}}{2^{3} 3^{17} \pi^{79}}\right)^{1/54}
%\left(\frac{T_n^4 \epsilon_L^4 \epsilon_M^{13}}{\mu_{\rm H_2}^{16} \mu_i^4}\right)^{1/9}
%\left(\frac{\alpha^{16} h^{24} c^{32}}{\Theta_c^{24} G^{27} \mh^{32}}\right)^{1/18} P^{-1/18}
 %\left(\frac{1.18^{64}}{2^{27} 3^{34} 5^{3} \pi^7}\right)^{1/54} \left(T_n^4 \epsilon_L^4 \epsilon_M^{13}\right)^{1/9} \left[\left(\frac{k_B}{\mu_{\rm H_2}}\right)^{32}\left(\frac{E_G}{\mu_i \sigma_{\rm SB}}\right)^{8}\frac{1}{\mh^{40} G^{27} \Theta_c^{24}}\right]^{1/18} P^{-1/18}
\\
& = & 0.15 \left(\frac{P/k_B}{10^6\,\rm K\,cm^{-3}}\right)^{-1/18}\,\msun,
\label{eq:mchapprox}
\end{eqnarray}
where $\alpha_G = G \mh^2/\hbar c=5.91\times 10^{-39}$ is the gravitational fine structure constant defined for two protons, $P_P = c^7/\hbar G^2 = 4.63\times 10^{114}$ dyn cm$^{-2}$ is the Planck pressure, and in the numerical evaluation in equation (\ref{eq:mchapprox}) I have used $\epsilon_L = 3/4$, $\epsilon_M=1/2$, $T_n = 1.86$, and $\Theta_c = 12.4$. Note that the dust abundance $\delta$ drops out of the problem entirely; numerical simulations show that this is an excellent approximation \citep{myers11a}.

\begin{figure}
\epsscale{1.18}
\plotone{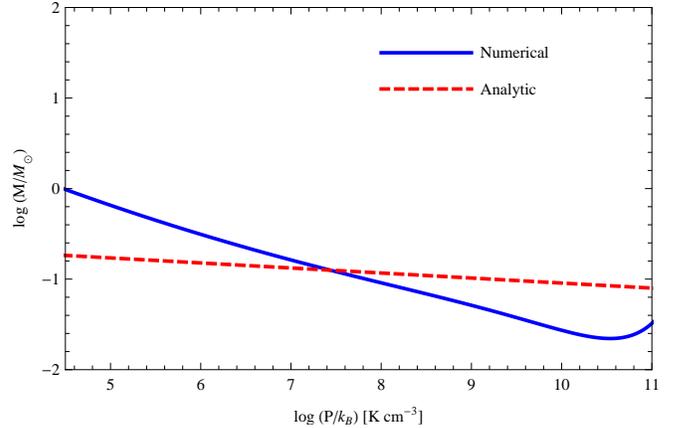}
\caption{
\label{mstar1}
Characteristic stellar mass $M$ as a function of interstellar pressure $P$, comparing the numerical solution (blue solid line) and the analytic approximation (red dashed line; equation (\ref{eq:mchapprox})). The numerical solution is for $\krho = 3/2$, $\delta = 1$, $\beta = 2$, $n=3/2$, $\epsilon_M = 1/2$, and $\epsilon_L = 3/4$. The slight upturn at very high pressures is associated with the dust photosphere moving past the outer edge of the thermal zone of influence around the star. Note that the vast majority of star-forming systems in the Galaxy lie at $P/k_B \ltsim 10^{8.5}$ K cm$^{-3}$, so higher $P/k_B$ values are realized only in extreme extragalactic environments, if at all.\\
}
\end{figure}

\begin{figure}
\epsscale{1.18}
\plotone{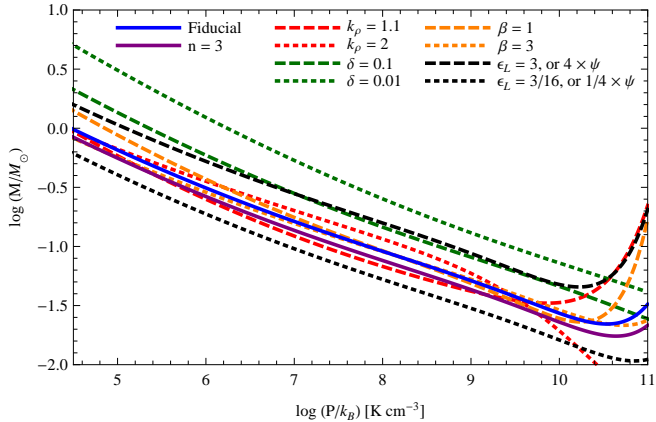}
\caption{
\label{mstar2}
Characteristic stellar mass $M$ as a function of interstellar pressure $P$, for varying parameters. The solid blue line is for the same parameters as in Figure \ref{mstar1}. The other lines show solutions in which one parameter is different: $n=3$ (solid purple), $\krho=1.1$ or $2$ (dashed and dotted red), $\delta=0.1$ or $0.01$ (green dashed and dotted), $\beta=1$ or $3$ (red dashed and dotted), and $\epsilon_L = 3$ or $3/16$ (equivalent to multiplying $\psi$ by 4 or $1/4$ relative to the fiducial estimate; black dashed and dotted). I use $\krho=1.1$ rather than $\krho = 1$ because formally the \citep{chakrabarti05a} approximation becomes singular at $\krho=1$; however, numerical solutions indicate that the results are nearly the same as for $\krho=1.1$. I do not show the results of varying the geometric parameter $\epsilon_M$ because this should not vary systematically with interstellar environment, and it simply provides an overall scaling. Note that, as for Figure \ref{mstar1}, the vast majority of Galactic star formation occurs at $P/k_B \ltsim 10^{8.5}$ K cm$^{-3}$. Also note that model values that fall below $0.01$ $\msun$ should not be taken seriously, since this below the estimated mass at which second collapse to stellar density occurs \citep{masunaga00a}.\\
}
\end{figure}

Pressures in star-forming systems cover a $4-6$ decade range, from the relatively diffuse molecular clouds found in nearby dwarf galaxies \citep{bolatto08a} (surface density $\Sigma \sim 0.01$ g cm$^{-2}$, corresponding to a pressure $P/k_B \sim G\Sigma^2/k_B \sim 3\times 10^4$ K cm$^{-3}$) to the densest star-forming gas clumps seen in the Galaxy ($\Sigma \sim 3$ g cm$^{-2}$, corresponding to $P/k_B \sim 3 \times 10^9$ K cm$^{-3}$). In extragalactic stellar systems such as the cores of giant ellipticals \citep{van-dokkum08b} and super star clusters \citep{turner00a} we see surface densities that reach even higher values of $\Sigma \sim 20$ g cm$^{-2}$. We do not know if these systems formed from gas at similarly high surface densities, but if they did the corresponding pressures would be $P/k_B \sim 10^{11}$ K cm$^{-3}$. Equation (\ref{eq:mstar}) predicts that even over this very large pressure range, the characteristic stellar mass should vary by only $\sim 1/3$ of a dex. For comparison, Figure \ref{mstar1} shows an exact numerical solution for $M_*$ as a function of pressure. We see that, while the pressure-dependence is slightly steeper than that predicted in the analytic approximation (mainly because $\kt$ is not exactly $0.5$), the characteristic mass still varies by only a decade or so as the pressure varies by more than six decades.

%For comparison, Figure \ref{mstar1} shows an exact numerical solution. We see that, while the pressure-dependence is slightly steeper than that predicted in the analytic approximation (mainly because $\kt$ is not exactly $0.5$), the characteristic mass still varies by only a decade over the more than six decade range that characterizes star-forming regions from the relatively diffuse molecular clouds found in nearby dwarf galaxies \citep{bolatto08a} (surface density $\Sigma \sim 0.01$ g cm$^{-2}$, corresponding to a pressure $P/k_B \sim G\Sigma^2/k_B \sim 3\times 10^4$ K cm$^{-3}$) to conditions found in the cores of giant ellipticals \citep{van-dokkum08b} or the densest super star-clusters \citep{turner00a} ($\Sigma \sim 20$ g cm$^{-2}$, corresponding to $P/k_B \sim 10^{11}$ K cm$^{-3}$).

Figure \ref{mstar2} shows numerical solutions for varying stellar polytropic indices, density powerlaw indices, metallicities, dust spectral indices, and
radiative energy budgets (i.e.\ values of $\epsilon_L$ or $\psi$)\footnote{The factor of $4$ variation is $\psi$ or $\epsilon_L$ is chosen to encompass the error in the value of $\psi$ that results from the polytropic approximation. This error shown in Figure \ref{fig:psiplot}b.}. Strikingly, most of these factors make no significant difference. Varying $n$, $k_\rho$, $\beta$, $\epsilon_L$, or $\psi$ within the plausible ranges of variation indicated by the upper and lower curves in the Figure produces less than a factor of 1.6 change in the characteristic stellar mass at all pressures $P/k_B<10^{10}$ K cm$^{-3}$. The only factor that matters marginally more is metallicity; decreasing $\delta$ to $0.1$, i.e.\ using a metallicity that is roughly $1/10$ the Solar value, induces a factor of 2 change in the characteristic mass, while using $\delta = 0.01$ produces a factor of $3-5$ shift. This confirms the analytic expectation that the properties of the interstellar environment -- metallicity, dust properties, and degree of gas concentration -- change the characteristic mass very weakly or not at all.

\section{Discussion and Conclusions}

The central results of this paper are equation (\ref{eq:mstar}) and figure \ref{mstar1}, which describe the characteristic stellar mass in terms of the hydrogen mass multiplied by a series of dimensionless factors. Some of these describe the geometry of the stellar accretion flow ($\epsilon_L$, $\epsilon_M$), the internal structure of protostars ($T_n$), and the chemical composition of gas ($\mu_{\rm H_2}$, $\mu_i$), and are always $\sim 1$. Others depend on the relative strength of electromagnetic, gravitational, and nuclear forces ($\alpha$, $\alpha_G$); these are fundamental constants. The result also depends on $\Theta_c$, which describes the energy scale in a stellar core in units of the Gamow energy. This is set mostly by the properties of the deuterium plus hydrogen fusion reaction, which also ultimately depends on fundamental constants. Finally, the last term depends on the interstellar pressure measured in units of the Planck pressure; this is the only term that makes any reference to interstellar conditions, and there with an extraordinarily weak dependence. We can therefore understand why the characteristic stellar mass should be invariant over such a broad range of conditions: it is set almost entirely by fundamental constants, with an almost vanishing dependence on interstellar conditions.

Furthermore, this result naturally explains why the stellar mass scale is such that nuclear reactions can be ignited in stars. Until deuterium burning begins in stellar cores, stars contract rapidly as they gain mass, their cores heat up, and $\psi$ becomes a strongly increasing function of mass. During this phase, as stars gain mass their thermal zone of influence rapidly expands, since increasing mass also increases the energy yield from accretion. Only once nuclear burning begins and the stellar core temperature is stabilized does the energy yield from accretion become roughly constant, and the zone of influence ceases to expand as rapidly, favoring fragmentation. Thus the onset of fragmentation is directly linked to stars reaching a mass such that nuclear reactions can begin.

Finally, I do find a very weak residual dependence of the characteristic stellar mass on the interstellar pressure and metallicity. These effects are small enough that they are likely to be masked within a single galaxy, or even over a wide range of galaxies of relatively similar properties, by random variations in factors like the accretion geometry, dust properties, and interstellar pressures. However, the dependence on pressure and metallicity may produce noticeable variations in samples that include galaxies where stars formed under conditions radically different than those found today. In particular, I find that the characteristic mass decreases weakly but noticeably in very high pressure and high metallicity environments such as the cores of giant elliptical galaxies. There is preliminary evidence for such a bottom-heavy IMF based on the presence of unexpectedly strong absorption features characteristic of very low mass stars in spectra taken from the central portions of giant ellipticals \citep{van-dokkum10a, van-dokkum11a}. At this point any link between this observational result and the theoretical one I derive here is necessarily speculative. We have no direct knowledge of the properties of the gas from which these stars formed, and it is possible that the pressure was less than one would infer from density of the final stellar system. Even if the pressures are high, we possess of a limited understanding of the physics of star formation in such extreme environments. Nonetheless, this work points to the need for further investigation of star formation at very high pressures, both observationally and theoretically.

\acknowledgements I acknowledge F.~Adams, C.~Conroy, C.~McKee, and P.~Schecter for discussions and comments on the manuscript,  T.~Hosokawa for providing the numerical stellar structure model shown in Figure \ref{fig:psiplot}, and an anonymous referee for helpful comments. This work was supported by the Alfred P.\ Sloan Foundation, NASA through ATFP grant NNX09AK31G, a Spitzer Space Telescope theoretical research program grant, and a Chandra Telescope grant, and the NSF through grants AST-0807739 and CAREER-0955300.

%\bibliographystyle{apj}
%\bibliography{refs}

\end{document}